\documentclass[floatfix,preprint,showpacs,nofootinbib]{revtex4}
\usepackage{natbib}
\usepackage{times}
\usepackage{amssymb,amsbsy,amsmath,amsfonts}
\usepackage{graphicx}
\usepackage{float}
\usepackage{morefloats}
\usepackage{rotating}
\newcommand{\eq}{\begin{eqnarray}}
\newcommand{\en}{\end{eqnarray}}
\begin{document}

\title{Two-photon and one photon-one vector meson decay widths of the $f_0(1370)$,
$f_2(1270)$, $f_0(1710)$, $f'_2(1525)$, and $K^*_2(1430)$}

\author{T. Branz,$^1$ L.S. Geng,$^2$ and E.~Oset$^2$}
\affiliation{$^1$Institut f\"ur Theoretische Physik,
Universit\"at T\"binged,
\\ Auf der Morgenstelle 14, D-72076 T\"binged, Germany}
\affiliation{$^2$Departamento de F\'{\i}sica Te\'orica and IFIC,
Centro Mixto Universidad de Valencia-CSIC,
Institutos de Investigaci\'on de Paterna, Apartado de Correos 22085, 46071 Valencia, Spain}

 \begin{abstract}
We calculate the radiative decay widths, two-photon ($\gamma\gamma$) and one photon-one vector meson ($V\gamma$), of the
dynamically generated resonances from vector meson-vector meson interaction in a unitary approach based on the hidden-gauge Lagrangians. In the present paper we consider the following dynamically generated resonances: $f_0(1370)$, $f_0(1710)$, $f_2(1270)$, $f_2'(1525)$, $K^*_2(1430)$, two
strangeness=0 and isospin=1 states, and two strangeness=1 and isospin=1/2 states.
For the $f_0(1370)$ and $f_2(1270)$ we reproduce the previous results for the
two-photon decay widths and further calculate their one photon-one vector decay widths.
For the $f_0(1710)$ and $f_2'(1525)$ the calculated two-photon decay widths are found to be consistent with data.
The $\rho^0\gamma$, $\omega\gamma$ and $\phi\gamma$ decay widths of the $f_0(1370)$, $f_2(1270)$, $f_0(1710)$, $f'_2(1525)$
are compared with the results predicted by other approaches. The $K^{*+}\gamma$ and $K^{*0}\gamma$ decay rates
of the $K^*_2(1430)$ are also calculated and compared with the results obtained in the framework of the covariant oscillator quark model.
The results for the two states with strangeness=0, isospin=1 and two states with strangeness=1, isospin=1/2
are predictions that need to be tested by future experiments.
\end{abstract}


\date{\today}

\maketitle

\section{Introduction}
One of the central topics in studies of low-energy strong
interaction is to understand how quarks and gluons combine into
hadronic objects that we observe experimentally, in other words, to understand low-energy
meson and baryon spectroscopy. Unfortunately, the
non-perturbative nature of QCD at low-energies has made a complete solution of
this problem from first principles almost impossible (admittedly,
lattice QCD has made remarkable progress in recent years,
and may provide a solution in the future). Furthermore, most of the
observed hadronic states are not asymptotic states, and as such,
they appear only in invariant mass distributions, phase shifts, etc.
This latter feature then implies that in many cases one can not ignore final
state interaction among their decay products.

A prominent example is the existence and nature of the $f_0(600)$.
For a comprehensive discussion and references, see the mini-review
``Note on scalar mesons'' of Ref.~\cite{Amsler:2008zzb}. Although
its existence has long been hypothesized, it took quite a long time
until different experiments have finally pinned it down unanimously.
Its nature is even more troubling, i.e., whether it is a genuine
$q\bar{q}$ state, $qq\bar{q}\bar{q}$ state, or molecular state. In
this context, the unitarization technique in combination with the
chiral Lagrangians, the so-called unitary chiral theories, have
provided a self-consistent picture where the $f_0(600)$ may be due
to the $\pi\pi$ final state
interactions~\cite{Oller:1997ti,Kaiser:1998fi,Markushin:2000a,Dobado:1997a,ramonet}.
The same approach has been used to study various other hadronic
systems, e.g., the kaon-nucleon
system~\cite{Kaiser:1995eg,Oset:1998a,ollerulf,Recio:2004a,Jido:2003a,Recio:2006a,Hyodo:2003a,Borasoy:2005a,Oller:2005a,Borasoy:2006a},
heavy-light systems~\cite{Kolomeitsev:2003ac} and three body
systems~\cite{MartinezTorres:2007sr}.

The unitary chiral approach, however, can only be employed to study interactions among
the Goldstone-bosons themselves and those between them and other
hadrons, because chiral symmetry only defines the interactions involving
the Goldstone-bosons. One may think about applying the same unitarization technique to study
other systems by employing phenomenological Lagrangians. In Refs.~\cite{Molina:2008jw,Gonzalez:2008pv,Sarkar:2009kx,Molina:2009eb,Oset:2009vf,Geng:2008gx},
by combining the phenomenologically successful hidden-gauge Lagrangians with
the above-mentioned unitarization technique, the interactions of vectors mesons
among themselves and with octet- and decuplet-baryons have been studied.
In the framework of this approach many interesting results have been obtained, which all
 compare rather favorably with
existing data.  The dynamically generated resonances
 should contain sizable meson-meson or meson-baryon components in
their wave-functions, thus qualifying as ``molecular states.''

Whether such a picture
is correct or partially correct has ultimately to be judged either
by data or by studies based on first principles (e.g., lattice QCD calculations). From the
first perspective, one should test as extensively as possible whether the proposed
picture is consistent with (all) existing data, make predictions, and propose experiments
where such predictions can be tested. These would provide further support to,
 or reject, the proposed nature of these states as being dynamically generated.

In the case of the vector meson--vector meson molecular states
obtained in Refs.~\cite{Molina:2008jw,Geng:2008gx}, several such
tests have been passed: In Refs.~\cite{Geng:2008gx,Geng:2009gb} it
has been shown that the branching ratios into
pseudoscalar-pseudoscalar and vector-vector final states of the
$f_0(1370)$,  $f_0(1710)$, $f_2(1270)$, $f_2'(1525)$, and
$K_2^*(1430)$ are all consistent with data. In Ref.~\cite{Nagahiro:2008um}, the two-photon
decay widths of the $f_0(1370)$ and $f_2(1270)$ have been calculated and found to agree with data.
Furthermore, in
Ref.~\cite{MartinezTorres:2009uk}, the ratios of the $J/\psi$ decay
rates into a vector meson ($\phi$, $\omega$, or $K^*$) and one of
the tensor states [$f_2(1270)$, $f_2'(1525)$, and $K^*_2(1430)$] have
been calculated, and the agreement with data is found to be quite reasonable.
Following the same approach, in Ref.~\cite{Geng:2009iw} it is shown that the ratio of the $J/\psi$ decay
rates into $\gamma f_2(1270)$ and $\gamma f'_2(1525)$ also agrees with data.

The radiative decay of a mesonic state has long been argued to be
crucial in determinations of the nature of the
state~\cite{Pennington:2008qa}. For instance, the non-observation of
the $f_0(1500)$ decaying into two photons has been used to support
its dominant glue nature~\cite{Amsler:2002ey}. In
Ref.~\cite{Nagahiro:2008um}, the two-photon decay widths of the
$f_0(1370)$ and $f_2(1270)$ have been calculated and found to agree
with data which therefore provides further support to the proposed
$\rho\rho$ molecular nature of these states~\cite{Molina:2008jw}. In
the present paper, we extend our previous work to the $f_0(1710)$, $f_2'(1525)$, $K^*_2(1430)$,
and four other states dynamically generated from vector meson -- vector meson interaction~\cite{Geng:2008gx}.
By taking into account all the SU(3) allowed coupled
channels, we also recalculate the two-photon decay widths of the
$f_0(1370)$ and $f_2(1270)$, which confirms the earlier results of
Ref.~\cite{Nagahiro:2008um} and provides a natural estimate of
inherent theoretical uncertainties.  We will also calculate the
one photon-one vector meson decay widths of these resonances.
As we will show below, in contrast to the results obtained in other theoretical models, our results show some
distinct patterns, which should allow one to distinguish between different models once data are available.

This paper is organized as follows: In Section 2, we explain in
detail how to calculate the two-photon and one photon-one vector meson decay widths of the
dynamically generated states. In Section 3, we compare the results with those
 obtained in other approaches and available data,
followed by a brief summary in Section 4.

\section{Formalism}
\subsection{Dynamically generated resonances from the vector meson-vector meson interaction}
In the following, we briefly
outline the main ingredients of the unitary approach (details can be found in Refs.~\cite{Molina:2008jw,Geng:2008gx}).
There are two basic building-blocks in this approach: transition amplitudes
provided by the hidden-gauge Lagrangians~\cite{Bando:1984ej} and a unitarization procedure. We adopt the Bethe-Salpeter equation method
 $T=(1-VG)^{-1}V$
 to unitarize the transition amplitudes $V$ for $s$-wave interactions,
where $G$ is a diagonal matrix
of the vector meson-vector meson one-loop function
\begin{equation}
 i\int\frac{d^4q}{(2\pi)^4}\frac{1}{q^2-M_1^2}\frac{1}{q^2-M_2^2}
\end{equation}
with $M_1$ and $M_2$ the masses of the two vector mesons.

\begin{figure}[t]
\centerline{\includegraphics[scale=0.35]{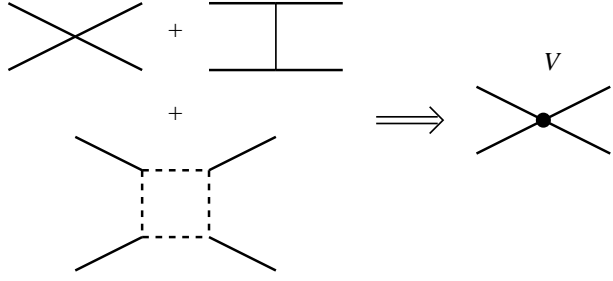}}
\caption{Transition amplitudes $V$ appearing in the coupled-channel Bethe-Salpeter equation.\label{fig:dia1}}
\end{figure}

In Refs.~\cite{Molina:2008jw,Geng:2008gx}
three mechanisms,  as shown in Fig.~\ref{fig:dia1}, have been taken into account for the transition amplitudes $V$:
the four-vector contact term, the t(u)-channel vector exchange amplitude, and the direct box amplitude with two intermediate pseudoscalar mesons.
Other possible mechanisms, e.g. s-channel vector exchange, crossed box amplitudes and
box amplitudes involving anomalous couplings, have been neglected,
since their contribution was found to be quite small in the detailed study
of  $\rho\rho$ scattering in Ref.~\cite{Molina:2008jw}.

Among the three mechanisms considered for $V$, the
four-vector contact term and $t(u)$-channel vector exchange one are responsible for
the formation of resonances or bound states provided that the interaction generated by them
is strong enough. In this sense,
the dynamically generated states can be thought of as ``vector meson-vector meson molecules.'' On the
other hand, the consideration of the imaginary part of the direct box amplitude allows the generated states to decay into two pseudoscalars. It should
be stressed that in the present approach these two mechanisms play quite different roles:
the four-vector contact interaction and the $t(u)$-channel vector exchange term are responsible
for generating the resonances or bound states, whereas the direct box amplitude mainly contributes to their decays. This particular
feature has an important consequence for the calculation of the radiative decay widths of the dynamically generated states
as shown below.

The one-loop function, Eq.~(1), is divergent and has to be
regularized. In Ref.~\cite{Geng:2008gx}, both dimensional
regularization method and cutoff method have been used. The
couplings of the dynamically generated states to their coupled
channels are given in Tables I, II, and III of
Ref.~\cite{Geng:2008gx}, which we need to calculate the radiative
decay widths of these resonances as explained below. In
Ref.~\cite{Geng:2008gx}, the couplings were obtained on the second
Riemann sheet using the dimensional regularization method without
including the box diagrams in the model. If, instead, the loop
functions were regularized using the cutoff method, one had to
calculate the couplings from the modulus of amplitude squared on the
real axis as done in Ref.~\cite{Nagahiro:2008um}. These two
approaches were found to yield consistent values for the couplings.
One has also some freedom in the values of the subtraction constants
due to data uncertainty and the coupled-channel nature of the
problem. An analysis of the resulting uncertainties has been
performed in Refs.~\cite{MartinezTorres:2009uk,Geng:2009iw}. They,
however, were found to translate into small uncertainties (at the
order of a few percent) in the present calculation.

We take advantage here to clarify a question often raised in
connection with the dynamically generated states. Since we all
accept that quarks are present in the physical mesons, the obvious
question is what happens to the ordinary $q\bar{q}$ states? The
answer to this can be found in the works of
Refs.~\cite{tornqvist1,tornqvist2,beveren,boglione}. In
those works, where the study of the scalar mesons is addressed,  one
starts with a seed of $q\bar{q}$ states representing scalar states
around $1.4$ GeV. Yet, these states unavoidably couple to meson
meson components. This is a necessity imposed by unitarity, since
the meson meson decay channels certainly couple to the physical
states. Invoking symmetries, like SU(3), other meson meson channels,
even those closed for the decay, will also couple to those $q
\bar{q}$ components. For instance the $f_0(980)$ resonance decays
into $\pi \pi$, so this must be a necessary coupled channel.
However, the underlying SU(3) symmetry of the strong interactions
will impose also the coupling to the $K \bar{K}$ component. One
rightly guesses that other channels with masses far away from that
of the $f_0(980)$ will play a minor role and can be neglected
(actually they can be accounted for, as we shall discuss below).
Then one has a coupled channel problem with $q \bar{q}$, $\pi \pi$
and $K \bar{K}$. According to Refs.~\cite{tornqvist1,tornqvist2,
beveren,boglione} the solution of the coupled channel problem leads
to the scalar states where the original $q \bar{q}$ states are
represented by a component of the wave function of minor importance,
since the meson meson cloud has taken over and represents the bulk
of the wave function. In simple words we can give a picture for this
situation. As is well known, when we give energy to a hadron to
break it and eventually see the quark components, we do not see the
quarks, we see mesons produced. This seems to be the case not only
when we break the hadron but when we excite it, such that the
creation of mesons becomes energetically more favorable that the
excitation of the quarks. One can easily visualize this in the
baryon spectrum: either the Roper or the N*(1535) resonances would
require 500-600 MeV of quark excitation energy, if they correspond
to genuine quark excitations. It is clear that the introduction of a
pion on top of the nucleon is energetically more favorable, so one
should investigate the pion nucleon dynamics (together with other
SU(3) related coupled channels) to see if this dynamics is able to
produce these states. Indeed, the N*(1535) appears as dynamically
generated from the meson baryon interaction in coupled
channels~\cite{kaiser,inoue}.

One can then still rightfully ask where the quark states go. Are
there within this picture states that are mostly of $q \bar{q}$
nature? The answer is yes in principle, but nothing can guarantee
it. One might  think that they should appear at higher
energies given the large energy needed to excite quarks. However,
this is not necessarily true as we shall comment at the end of this
section. On the other hand, the meson meson channels of smaller
energy will be open. This detail should not go unnoticed. Indeed,
let us think of a single channel problem with an attractive
potential. One can get many discrete bound states in principle. Let
us add another channel with an attractive potential, which by itself
also generates discrete bound states. When we allow some coupling
among these two channels then the earlier initial states give rise
to two orthogonal combinations of the two channels. One might expect
the same thing when we put together meson and quark channels. Yet,
the counting of states does not follow here because for higher
energies the meson meson channel will be unbound and then we can
have a continuum of states. We can of course find out resonances,
but this is not guaranteed nor is there any rule on how many
resonances should appear. It all depends on the dynamics. The
problem is indeed very interesting, but as far as one restricts
oneself to low-lying resonances the meson meson nature is prominent
and the effective Lagrangians used to take care of their interaction
lead naturally to some bound states, which are those we consider. As
to whether there are other states of simpler quark nature, in our
approach we cannot say anything since these components are not part
of our coupled channels states. However, apart from the works
mentioned earlier~\cite{tornqvist1,tornqvist2,beveren} there are
works in this direction in Refs.~\cite{vijande,vijande2}, which also
conclude that the states of lower energy are mostly of mesonic
nature.

Continuing with these observations, in connection with the quark
components one can say that even the small admixture of these quark
components could change the mass and other properties of the
resonances. This might be so to some extent, but the studies with
chiral dynamics and only hadron components have an element in the
formalism which allows one to take this into account in an effective
way. This is the subtraction constant in the $G$ function when the
dimensional regularization formula is used, or the cut off in the
cut off method. The basic idea of having the hadronic components as
main building blocks is that the spectra is obtained using a natural
value for the cut off or the subtraction constant~\cite{ollerulf}.
Fine tuning of these subtraction constant or cut off can take into
account the contribution from additional channels not explicitly
considered in the approach, like the quark states~\cite{hyodo,bugg}.
In fact sometimes one needs a massive change of the cut off to
reproduce the mass of a particle, which is a clear manifestation
that the state under consideration is not of hadronic, but more of
quark nature. This is the case for the $\rho$ meson, which does not
come as an object made of $\pi \pi$. In the study of $\pi \pi$
scattering using the lowest-order chiral Lagrangians it would
require a cut off of the order of several TeV, which is obviously
far away from the natural scale of 1 GeV  in effective theories of
the low energy hadron spectra~\cite{ramonet,jose1,jose2}.

Actually the case of the $\rho$ is a good example of warning
concerning the dynamical generation of resonances.  If in the $\pi
\pi$ interaction one takes the leading- and next-to-leading-order of
the $s$-channel $\rho$-exchange amplitude and unitarizes it with the
IAM (inverse amplitude method) or the Bethe Salpeter equation, one
obtains the full amplitude (see section III of Ref.~\cite{ramonet}).
This is further elaborated in Ref.~\cite{giacosa1}, which warns that
this can happen in unitarization procedures, inducing one to think
that one obtains a dynamically generated resonance, when in fact one
is merely regenerating a preexisting resonance, which has been
integrated out of the original Lagrangian and is not contained in
the effective Lagrangian as a fundamental field. Although this
warning should be kept in mind, one should also note that apart from
regenerating a preexisting resonance, one can, and does in practice,
generate other non-preexisting  ones due to other terms in the
potential, different from those directly associated to the
$s$-channel exchange of the preexisting resonance, like contact
terms and $t$- and $u$-channel exchange of those preexisting
resonances. This is particularly clear in the case of the low-lying
scalar mesons, which have different quantum numbers than the $\rho$.
The latter,  as mentioned above, would be ``regenerated" in the
unitarization scheme using the leading- and next-to-leading-order
terms in the potential.

  Nevertheless, in spite of all the arguments given in favor of the dynamically
   generated vector-vector states, the fact remains that the tensor states $f_2(1270)$,
   $f'_2(1525)$, $a_2(1320)$, $K^*_2(1430)$ are well reproduced in the quark model,
   including many of their decay modes (see, e.g., Ref.~\cite{Li:2000zb,klempt,crede,isgur,Barnes:1996ff,Barnes:2002mu,Anisovich:2002im}).
   This success in both models may reflect the fact that the constituent quarks in
   quark models are objects effectively dressed with meson clouds and the overlap between
   the molecular picture and the quark model picture could be bigger than expected in some cases~\cite{gonzalez}.
   Yet, even in this case, using one picture or the other could be more suited for other observables than those
   where the two models succeed. It is thus worth working with both models to make predictions.
   As we shall see in Section III, there are some observables where the predictions of the two models
   are indeed rather different.

\subsection{Radiative decays, $\gamma\gamma$ and $V\gamma$, of the dynamically generated resonances}
A detailed explanation of the two-photon decay mechanism has been given in Ref.~\cite{Nagahiro:2008um}.
Here, we follow closely Ref.~\cite{Nagahiro:2008um} and extend it to the case of one photon-one vector meson decay.

\begin{figure}[t]
\centerline{\includegraphics[scale=0.6]{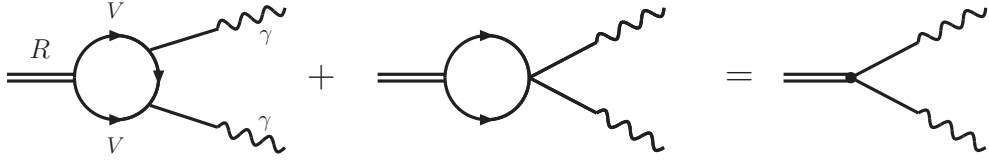}}
\caption{Two-photon decay of a dynamically generated resonance from vector meson-vector meson interaction.\label{fig:twogamma}}
\end{figure}

The coupling of a photon to a dynamically generated resonance goes through
couplings to its coupled-channel components in all possible ways such that
gauge invariance is conserved (see, e.g, Refs.~\cite{Nacher:1999ni,Borasoy:2007ku} for a relevant discussion
within the kaon-nucleon system).
A peculiar feature of the hidden-gauge Lagrangians is that photons do not
couple directly to charged vector mesons but indirectly through their conversion to
$\rho^0$, $\omega$, and $\phi$. This, together with the fact that
the four-vector contact and the t(u)-channel exchange diagrams are responsible
for the generation of the resonances or bound states, imply that the coupling of a photon to
the resonance (or bound state) can be factorized into a strong part and an electromagnetic part~\cite{Nagahiro:2008um}\footnote{We refer to the same
reference for a demonstration of gauge invariance of this approach.}:
i.e., the resonance first decays into two vector mesons and then one
or both of them convert into a photon. This is demonstrated schematically in Fig.~\ref{fig:twogamma} for the case
of the two-photon decay. In the case of one photon-one vector meson decay, one simply replaces one of the final photons by one
vector meson.
\begin{figure}[t]
\centerline{\includegraphics[scale=0.75]{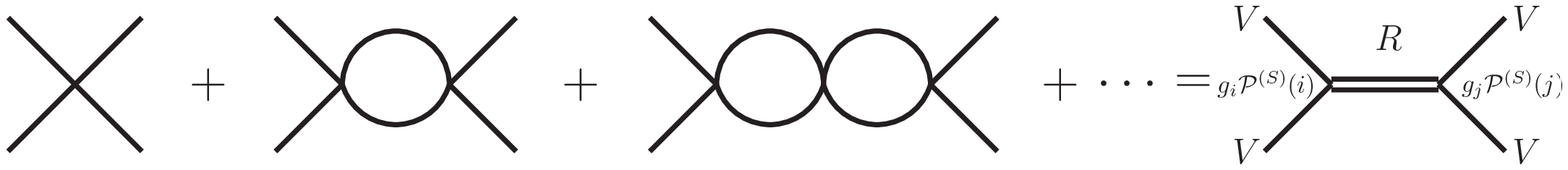}}
\caption{Pole representation of the vector-vector scattering amplitude and the definition of couplings of a dynamically
generated resonance to its components.\label{fig:rescoup}}
\end{figure}

Close to a pole position, the vector-vector scattering amplitude given in Fig.~\ref{fig:rescoup} can be parameterized
as
\begin{equation}
 T_{ij}^{(S)}=g_i \mathcal{P}^{(S)}(i) \frac{1}{s-M_R^2+i M_R \Gamma_R} g_j\mathcal{P}^{(S)}(j)
\end{equation}
where $\mathcal{P}^{(S)}$  is the spin projection operator, which
projects the initial (final) vector meson-vector meson pair $i$ ($j$) into spin $S$ with
\begin{equation}\label{eq:proj0}
 \mathcal{P}^{(0)}=\frac{1}{\sqrt{3}}\epsilon_i(1)\epsilon_i(2),
\end{equation}
\begin{equation}\label{eq:proj1}
 \mathcal{P}^{(1)}=\frac{1}{2}\big[\epsilon_i(1)\epsilon_j(2)-\epsilon_j(1)\epsilon_i(2)\big],
\end{equation}
\begin{equation}\label{eq:proj2}
 \mathcal{P}^{(2)}=\frac{1}{2}\big[\epsilon_i(1)\epsilon_j(2)+\epsilon_j(1)\epsilon_i(2)\big]-
\frac{1}{3}\epsilon_m(1)\epsilon_m(2)\delta_{ij},
\end{equation}
where $\epsilon(1)$ [$\epsilon(2)$] is the polarization vector of particle 1 [2] and $i$, $j$, $m$ runs
from 1 to 3 since in line with the approximation made in Refs.~\cite{Molina:2008jw,Geng:2008gx}
that $|\vec{p}|\,/M_V$ is small and hence $\epsilon_0=0$.
The couplings $g_i$ ($g_j$) are obtained from the resonance pole position on the complex plane
and are tabulated in Ref.~\cite{Geng:2008gx}.\footnote{They can also be obtained from the study of the
transition amplitudes in the real axis as done in Ref.~\cite{Nagahiro:2008um}, where box diagrams can also be taken into account. We find
that differences between the couplings obtained in these two ways are very small
for the $f_0(1370)$, $f_0(1710)$, $f_2(1275)$, $f'_2(1525)$, and $K^*_2(1430)$, well within
the uncertainties that we estimate for the quantities we calculate in this work, $\sim20\%$.} To evaluate the two-photon and one photon-one vector partial
decay widths of the dynamically generated particles, one needs its coupling to the vector-vector components, i.e.,
$g_i \mathcal{P}^{(S)}_i$.

The amplitude of a neutral non-strange vector meson converting into a photon
is given by
\begin{equation}
 t_{V\gamma}= \mathcal{C}_{V\gamma}\frac{e}{g} M_V^2\epsilon^\mu(V)\epsilon_\mu(\gamma)\;\mbox{with}\;\mathcal{C}_{V\gamma}=\left\{\begin{array}{l}
                                                                  \frac{1}{\sqrt{2}}\;\mbox{for $\rho^0$}\\
                                                                  \frac{1}{3\sqrt{2}}\;\mbox{for $\omega$}\\
                                                                  -\frac{1}{3}\;\mbox{for $\phi$}
                                                                 \end{array}\right.,
\end{equation}
with $g=\frac{m_\rho}{2 f_\pi}$.
Therefore, the whole two-photon and one photon-one vector decay amplitudes for a resonance $R$ of
spin $S$ are
\begin{equation}\label{eq:trprop1}
 T^{(R)}(\gamma\gamma)\propto \sum\limits_{V_1,V_2} g^{(R)}_{V_1V_2} {\cal P}^{(S)}_{V_1 V_2}\left(\frac{1}{-M_{V_1}^2}\right)t_{V_1 \gamma}
\left(\frac{1}{-M_{V_2}^2}\right)t_{V_2 \gamma}\times F_1,
\end{equation}
\begin{equation}\label{eq:trprop2}
 T^{(R)}(V_1\gamma)\propto \sum\limits_{V_2} g^{(R)}_{V_1V_2} {\cal P}^{(S)}_{V_1 V_2}
\left(\frac{1}{-M_{V_2}^2}\right)t_{V_2 \gamma}\times F_1,
\end{equation}
where $F_1$ is a proper isospin coefficient which projects the vector-vector pair
in isospin space to that in physical space and $g_{V_1V_2}^{(R)}$ denotes the coupling of
resonance $R$ to channel $V_1 V_2$. Recall that in Ref.~\cite{Geng:2008gx} we
have used the following phase conventions: $K^{*-}=-|1/2,-1/2
\rangle$ and $\rho^+=-|1,+1\rangle$, which implies that
\begin{eqnarray}\label{eq:isospin}
 |\rho\rho\rangle_{I=0}&=&-\frac{1}{\sqrt3}|\rho^+\rho^-+\rho^-\rho^++\rho^0\rho^0\rangle,\nonumber\\
|\rho K^*\rangle_{I=1/2,I_3=1/2}&=&-\sqrt{\frac23}|\rho^+K^{*0}\rangle-\sqrt{\frac13}|\rho^0K^{*+}\rangle,\\
|\rho K^*\rangle_{I=1/2,I_3=-1/2}&=&\sqrt{\frac13}|\rho^0 K^{*0}\rangle-\sqrt{\frac23}|\rho^- K^{*+}\rangle\nonumber.
\end{eqnarray}
From Eq.~(\ref{eq:isospin}), one can easily read off the isospin projector $F_1$.

Summing over polarization of the intermediate vector mesons in Eqs.~(\ref{eq:trprop1},\ref{eq:trprop2}) and taking
into account symmetry factors and proper normalization, one has the following amplitudes
\begin{eqnarray}
T^{(R)}_{\gamma\gamma}&=&\frac{e^2}{g^2}\sum\limits_{V_1,V_2=\rho^0,\omega,\phi}g^{(R)}_{V_1V_2}
\mathcal{P}^{(S)}_{\gamma\gamma}{\cal C}_{V_1\gamma}
{\cal C}_{V_2\gamma}\times F_1 \times F_2\\
T^{(R)}_{V_1\gamma}&=&\frac{e}{g}\sum\limits_{V_2=\rho^0,\omega,\phi}g^{(R)}_{V_1V_2}\mathcal{P}^{(S)}_{V_1\gamma}
{\cal C}_{V_2\gamma}\times F_1 \times F_3
\end{eqnarray}
where $\mathcal{P}^{(S)}_{V\gamma}$  and  $\mathcal{P}^{(S)}_{\gamma\gamma}$ are defined
in Eqs.~(\ref{eq:proj0},\ref{eq:proj1},\ref{eq:proj2}) with $V$ ($\gamma$) denoting the polarization vector
of a vector-meson (photon);
 $F_1$ is the isospin factor, and $F_2$, $F_3$
account for both a symmetry factor and the unitary normalization used in Refs.~\cite{Molina:2008jw,Geng:2008gx}
\begin{equation}
 F_2 =\Bigg\{\begin{array}{ll}\sqrt{2}&\text{ for a pair of identical particles, e.g. $\rho^0\rho^0$}\\
2&\text{ for a pair of different particles, e.g. $\rho^0 \omega$}\end{array},
\end{equation}
\begin{equation}
 F_3 =\Bigg\{\begin{array}{ll}\sqrt{2}&\text{ for a pair of identical particles, e.g. $\rho^0\rho^0$}\\
1&\text{ for a pair of different particles, e.g. $\rho^0 \omega$}\end{array}.
\end{equation}

The two-photon and one photon-one vector decay widths of a dynamically generated resonance
$R$ of spin $S$ are then given by
\begin{eqnarray}
\Gamma_{\gamma\gamma}&=&\frac{1}{2S+1}\frac{1}{16\pi M_R}\frac{1}{2}\times
\sum\limits_{\mbox{polarization}}|T^{(R)}_{\gamma\gamma}|^2,\\
  \Gamma_{V\gamma }&=&\frac{1}{2S+1}\frac{1}{8\pi M_R}\frac{|p_\gamma|}{M_R}\times
\sum\limits_{\mbox{polarization}}|T^{(R)}_{V\gamma}|^2,
\end{eqnarray}
where $M_R$ is the resonance mass and $p_\gamma$ is the photon momentum in
the rest frame of the resonance $R$.
For the photon, we work in the Coulomb gauge ($\epsilon^0=0$ and $\vec{k}\cdot\vec{\epsilon}=0$), where the sum over the final polarizations are given by
\begin{equation}\label{pollambda}
 \sum_{\mbox{polarization}}\epsilon_i(\gamma)\epsilon^*_j(\gamma)=\delta_{ij}-\frac{k_i k_j}{|\vec{k}|^2}
\end{equation}
with $\vec{k}$ the three momentum of the photon. For vector mesons, one has $\epsilon_0=0$ [see discussion
below Eq.~(\ref{eq:proj2})] and
\begin{equation}\label{polvector}
 \sum_{\mbox{polarization}}\epsilon_i(V)\epsilon^*_j(V)=\delta_{ij}.
\end{equation}
With Eqs.~(\ref{pollambda},\ref{polvector}), one can easily verify
\begin{equation}
 \sum\limits_{\mbox{polarization}}{\cal P}^{(S)}_{\gamma\gamma}{\cal P}^{*(S)}_{\gamma\gamma}=
\left\{\begin{array}{ll}
\frac{2}{3}\quad& S=0\\
1\quad&S=1\\
\frac{7}{3}\quad&S=2
\end{array}\right.,
\end{equation}
\begin{equation}
 \sum\limits_{\mbox{polarization}}{\cal P}^{(S)}_{V\gamma}{\cal P}^{*(S)}_{V\gamma}=
\left\{\begin{array}{ll}
\frac{2}{3}\quad& S=0\\
2\quad&S=1\\
\frac{10}{3}\quad&S=2
\end{array}\right..
\end{equation}

\begin{table*}[t]
      \renewcommand{\arraystretch}{1.0}
     \setlength{\tabcolsep}{0.1cm}
     \centering
     \caption{Pole positions (in units of MeV) and radiative decay widths (in units of keV) in the strangeness=0 and isospin=0 channel.
     For the sake of reference, we also show the mass and width of the dynamically generated resonances
obtained taking into account the box diagrams with $\Lambda=1$ GeV
and $\Lambda_b=1.4$ GeV~\cite{Geng:2008gx}.
            \label{table:res1}}
     \begin{tabular}{cccccccc}
     \hline\hline
Pole position & (Mass, Width)& Meson&$\Gamma_{\rho^0\gamma}$&
$\Gamma_{\omega\gamma}$ & $\Gamma_{\phi\gamma}$ &
$\Gamma_{\gamma\gamma}$ & $\Gamma_{\gamma\gamma}$ (Exp.)\\\hline
$(1512,-i26)$ & $(1523,257)$  & $f_0(1370)$&726&0.04&0.01&1.31&-\\
$(1726,-i14)$ & $(1721,133)$ &$f_0(1710)$&24&82&94&0.05&$<0.289$
~\cite{Amsler:2008zzb}\footnote{ This rate is obtained using
$\Gamma_{\gamma\gamma}\times
\Gamma_{K\bar{K}}/\Gamma_\mathrm{total}<0.11$
keV~\cite{Behrend:1988hw} and
$\Gamma_{K\bar{K}}/\Gamma_\mathrm{total}=0.38^{+0.09}_{-0.19}$~\cite{Longacre:1986fh}.
On the other hand, if one uses
$\Gamma_{K\bar{K}}/\Gamma_\mathrm{total}\approx0.55$ obtained in
Ref.~\cite{Geng:2009gb}, one would obtain
$\Gamma_{\gamma\gamma}<0.2$ keV.}\\
$(1275,-i1)$ & $(1276,97)$ & $f_2(1270)$&1367&5.6&5.0&2.25&$3.03\pm0.35$~\cite{Amsler:2008zzb}\\
&&&&&&&$2.27\pm0.47\pm0.11$ \cite{Adachi:1989dd}\\
&&&&&&&$2.35\pm0.65$ \cite{Morgan:1990kw}\\
$(1525,-i3)$& $(1525,45)$ & $f_2^\prime(1525)$&72&224&286&0.05&$0.081\pm0.009$~\cite{Amsler:2008zzb}\\
\hline\hline
    \end{tabular} 
\end{table*}

\begin{table*}[htpb]
      \renewcommand{\arraystretch}{1.6}
     \setlength{\tabcolsep}{0.4cm}
     \centering
     \caption{The same as Table \ref{table:res1}, but for the strangeness=0 and isospin=1 channel.\label{table:res2}}
     \begin{tabular}{cccccccc}
     \hline\hline
Pole position&(Mass,Width) &Meson&$\Gamma_{\rho^0\gamma}$
&$\Gamma_{\omega\gamma}$  & $\Gamma_{\phi\gamma}$
&$\Gamma_{\gamma\gamma}$  \\\hline
$(1780,-i66)$ & $(1777,148)$ &$a_0$&247&290&376&1.61\\
$(1569,-i16)$ & $(1567,47)$ &$a_2$&327&358&477&1.60\\
 \hline\hline
    \end{tabular} 
\end{table*}
 \begin{table*}[htpb]
      \renewcommand{\arraystretch}{1.6}
     \setlength{\tabcolsep}{0.5cm}
     \centering
     \caption{The same as Table \ref{table:res1}, but for the strangeness=1 and isospin=1/2 channel.\label{table:res3}}
     \begin{tabular}{ccccc}
     \hline\hline
     Pole position &(Mass,Width)  & Meson&$\Gamma_{K^{\ast +} \gamma}$ & $\Gamma_{K^{\ast 0}\gamma}$ \\\hline
$(1643,-i24)$ & $(1639,139)$ &$K_0^\ast$& 187 & 520\\
$(1737,-i82)$ & $(1743,126)$ &$K_1$ &143      & 571\\
$(1431,-i1)$ &  $(1431,56)$ &$K_2^\ast(1430)$&261 &1056\\\hline
\hline
    \end{tabular} 
       \end{table*}
\section{Results and discussions\label{sec:res}}
       In this section, we discuss our main results and compare them with
available data and the predictions of other approaches. In Tables \ref{table:res1}, \ref{table:res2}, \ref{table:res3}, we
show the calculated one photon-one vector meson and two-photon decay widths of the
resonances dynamically generated in Ref.~\cite{Geng:2008gx}. We have also listed relevant data for the two-photon decay
widths from different experiments. It should be pointed out in our approach that among the 11 dynamically generated resonances~\cite{Geng:2008gx},
the $h_1$ state does not decay into $\gamma\gamma$ and $V\gamma$;
the same is true for the $b_1$ state; on the other hand, the $K_0^*$, $K_1$ and $K^*_2(1430)$ resonances only
decay into $K^*\gamma$ but not $\gamma\gamma$.

For the $f_2(1270)$ and $f_0(1370)$, Nagahiro et al. have calculated the two-photon decay widths
as 2.6 keV and 1.62 keV~\cite{Nagahiro:2008um}.  Recall in that work among all the SU(3) allowed channels only the $\rho\rho$ channel
was considered and also the couplings deduced from amplitudes on the real-axis were used. Therefore,
 the differences between
the two-photon decay widths obtained in the present work and those obtained in Ref.~\cite{Nagahiro:2008um} can
be viewed as inherent theoretical uncertainties, which are $\sim20\%$.  As also discussed in
Ref.~\cite{Nagahiro:2008um}, it is clear  from Table \ref{table:res1}
that our two-photon decay width for the $f_2(1270)$ agrees well with the data.  The experimental
situation for the $f_0(1370)$ is not yet clear, but as discussed in Ref.~\cite{Nagahiro:2008um}, current
experimental results are consistent with our result for $\Gamma_{\gamma\gamma}$.

In addition to the $f_2(1270)$ and $f_0(1370)$ we have
calculated the two-photon decay widths of the $f_0(1710)$ and $f'_2(1525)$. From
Table \ref{table:res1}, it can be seen that they agree reasonably well
with available data. Our calculated two-photon decay width for the $f'_2(1525)$ is slightly smaller than the experimental value
quoted in the PDG review. This is quite acceptable since 1)
as discussed earlier we have an inherent theoretical uncertainty of $\sim20\%$ and
2) there might be other relevant coupled channels that have not been taken into account
in the model of Ref.~\cite{Geng:2008gx}, which can be inferred from the fact that
the total decay width of the $f'_2(1525)$ in that model $\sim50$ MeV is smaller than the experimental value $\sim70$ MeV.

Note that the significantly small value of the widths of the $f_0(1710)$ and $f'_2(1525)$ compared to
that of the $f_2(1270)$, for example, has a natural interpretation in our theoretical framework since
the former two resonances are mostly $K^*\bar{K}^*$ molecules and therefore the couplings to $\rho\rho$, $\omega\omega$, $\omega\phi$, $\phi\phi$,
which lead to the final $\gamma\gamma$ decay, are very small. The advantages of working with coupled channels become obvious
in the case of these radiative decays. While a pure $K^\ast \bar{K}^*$ assignment would lead to $\Gamma_{\gamma\gamma}$=0 keV, our coupled channel analysis gives the right strength for the couplings to the weakly coupled channels.

In the following we shall have a closer look at the radiative decay widths of the $f_2(1270)$, $f'_2(1525)$,
$f_0(1370)$, $f_0(1710)$, $K^*_2(1430)$, and compare them with the predictions from other theoretical approaches.

\subsection{Radiative decay widths of $f_2(1270)$ and $f'_2(1525)$}

In Table \ref{table:f2comparison}, we compare our results for the radiative decay widths for the $f_2(1270)$ with those obtained in
other approaches, including the covariant oscillator quark model (COQM)~\cite{Ishida:1988uw},
the tensor-meson dominance (TMD)
model~\cite{Suzuki:1993zs}, the AdS/QCD
calculation in~\cite{Katz:2005ir}, the model assuming both tensor-meson
dominance and vector-meson dominance (TMD\&VMD)~\cite{Oh:2003aw},
and the nonrelativistic quark-model (NRQM)~\cite{Close:2002sf}. From this
comparison, one can see that the AdS/QCD calculation and our present
study provide a two-photon decay width consistent with the data. The
TMD model result is also consistent with the data (it can use either
$\Gamma(f_2(1270)\rightarrow\gamma\gamma)$ or
$\Gamma(f_2(1270)\rightarrow\pi^+\pi^-)$ as an input to fix its single parameter), while the TMD\&VMD model prediction is off by a factor of 3.
Particularly interesting is the fact that although the TMD\&VMD
model predicts $\Gamma(f_2(1270)\rightarrow\rho\gamma)$ similar to
our prediction, but in contrast their result for $\Gamma(f_2(1270)\rightarrow\omega\gamma)$ is much larger than ours,
almost a factor of 30. Therefore, an experimental measurement of the
ratio of
$\Gamma(f_2(1270)\rightarrow\rho\gamma)/\Gamma(f_2(1270)\rightarrow\omega\gamma)$
will be very useful to disentangle these two pictures of the $f_2(1270)$. Furthermore, one notices that
all theoretical approaches predict $\Gamma_{\rho^0\gamma}$ to be of the order of a few 100 keV.

In Table \ref{table:f2pcomparison}, we compare the radiative decay widths of the $f'_2(1525)$ predicted
in the present work with those obtained in the COQM~\cite{Ishida:1988uw}.  We notice that
the COQM predicts $\Gamma_{\phi\gamma}/\Gamma_{\rho^0\gamma}\approx22$
while our model gives an estimate of $\Gamma_{\phi\gamma}/\Gamma_{\rho^0\gamma}
\approx4$, which are quite distinct even taking into account model uncertainties.  Furthermore, $\Gamma_{\omega\gamma}$ in
the COQM is almost zero while it is comparable to $\Gamma_{\phi\gamma}$ in our approach. An experimental
measurement of any two of the three decay widths will be able to confirm either the COQM
picture or the dynamical picture.

\begin{table*}[t]
      \renewcommand{\arraystretch}{1.6}
     \setlength{\tabcolsep}{0.1cm}
     \centering
     \caption{Radiative decay widths of the $f_2(1270)$ obtained
     in the present work in comparison with those obtained in other approaches.\label{table:f2comparison}.}
         \begin{tabular}{ccccccc}
     \hline\hline
&COQM~\cite{Ishida:1988uw}&TMD~\cite{Suzuki:1993zs}\footnote{The model only provides ratios of
the $f_2(1270)$ decay rates. Therefore, if using the then quoted
experimental decay rate
$\Gamma(f_2(1270)\rightarrow\gamma\gamma)=3.15\pm0.04\pm0.39$
keV~\cite{Boyer:1990vu}, the model predicts
$\Gamma(f_2(1270)\rightarrow\rho\gamma)=630\pm86$ keV.}
&AdS/QCD\cite{Katz:2005ir}&
TMD\&VMD~\cite{Oh:2003aw}&NRQM~\cite{Close:2002sf}&Present
work\\\hline
$f_2(1270)\to\gamma \gamma$&- &$3.15\pm0.04\pm0.39$&2.71&8.8&-&2.25\\
$f_2(1270)\to\rho^0 \gamma$&254 & $630\pm86$&-&1364&644 &1367\\
$f_2(1270)\to \omega \gamma$&27&-&-&$167.6\pm25$& &5.6\\
$f_2(1270)\to\phi\gamma$ &1.3 &- & - & - & - & 5.0\\
\hline\hline
    \end{tabular} 
\end{table*}

\begin{table*}[t]
      \renewcommand{\arraystretch}{1.6}
     \setlength{\tabcolsep}{0.3cm}
     \centering
     \caption{Radiative decay widths of the $f'_2(1525)$ obtained
     in the present work in comparison with those obtained in the covariant oscillator quark model (COQM)~\cite{Ishida:1988uw}.\label{table:f2pcomparison}.}
         \begin{tabular}{cccc}
     \hline\hline
&COQM~\cite{Ishida:1988uw} &Present
work\\\hline
$f'_2(1525)\to\gamma \gamma$& &0.05\\
$f'_2(1525)\to\rho^0 \gamma$&4.8 &72\\
$f'_2(1525)\to \omega \gamma$&0&224\\
$f'_2(1525)\to\phi\gamma$ &104& 286\\
\hline\hline
    \end{tabular} 
\end{table*}

\begin{table*}[t]
      \renewcommand{\arraystretch}{1.6}
     \setlength{\tabcolsep}{0.4cm}
     \centering
     \caption{Branching ratio of $\Gamma(f_2^\prime(1525)\to\gamma\gamma)$ and $\Gamma(f_2(1270) \to\gamma\gamma)$
in comparison with that obtained in other approaches and
data.\label{table:ratio2f2}}
     \begin{tabular}{ccccc}
     \hline\hline
&EF~\cite{Giacosa:2005bw}&TMS~\cite{Li:2000zb}& PDG~\cite{Amsler:2008zzb}
&Present work\\\hline
$\Gamma(f_2^\prime(1525)\to\gamma\gamma)/\Gamma(f_2(1270) \to\gamma\gamma)$&0.046&0.034&$0.027\pm0.006$&0.023\\
\hline\hline
    \end{tabular}
\end{table*}

An interesting quantity in this context is the ratio
$\frac{\Gamma(f'_2(1525)\rightarrow\gamma\gamma)}{\Gamma(f_2(1270)\rightarrow
\gamma\gamma)}$ since naturally branching ratios suffer less from systematic uncertainties within a
model. In Table \ref{table:ratio2f2}, we compare our
result with data and those obtained in other approaches. It is clear
that our result lies within the experimental bounds while those of the effective field approach (EF)
~\cite{Giacosa:2005bw} and the two-state mixing scheme (TMS)~\cite{Li:2000zb} are slightly larger than the
experimental upper limit, with the latter being almost at the upper limit. Given the fact that we have no free
parameters in this calculation, such an agreement is reasonable.

\begin{table*}[htpb]
      \renewcommand{\arraystretch}{1.4}
     \setlength{\tabcolsep}{0.13cm}
     \centering
     \caption{Radiative decay widths of the $f_0(1370)$ and $f_0(1710)$ obtained in
     the present work in comparison with those obtained in other approaches.
     All decay widths are given in keV.}\label{table:f0comparison}
     \begin{tabular}{ccccccccccc}
     \hline\hline
&\multicolumn{3}{c}{NRQM~\cite{Close:2002sf}\footnote{Light, medium
and heavy indicate three possibilities for the bare glueball mass:
lighter than the bare $n\bar{n}$ state (Light), between that of the
bare $n\bar{n}$ state and that of the bare $s\bar{s}$ state
(Medium), and heavier than that of the bare $s\bar{s}$ state
(Heavy).}} &\multicolumn{3}{c}{LFQM~\cite{DeWitt:2003rs}$^a$}
&\multicolumn{2}{c}{\cite{Nagahiro:2008bn}}&\cite{Giacosa:2005zt}&Present\\
&Light& Medium &Heavy&Light& Medium& Heavy&$K\bar K$ loop&$\pi\pi$
loop&&work\\\hline
$f_0(1370)\to\gamma\gamma$&-& -& -&1.6&$3.9^{+0.8}_{-0.7}$&  $5.6^{+1.4}_{-1.3}$&-&-&0.35&1.31\\
$f_0(1370)\to\rho^0\gamma$&443&1121& 1540& 150&$390^{+80}_{-70}$&$530^{+120}_{-110}$&$79\pm40$&$125\pm80$&-&726\\
$f_0(1370)\to\omega\gamma$&-&-&-&-&-&-&$7\pm3$&$128\pm80$&-&0.04\\
$f_0(1370)\to\phi\gamma$& 8&9&32 &0.98&$0.83^{+0.27}_{-0.23}$&$4.5^{+4.5}_{-3.0}$&$11\pm6$&-&-&0.01\\
\hline
$f_0(1710)\to\gamma\gamma$&- &-& -&0.92&$1.3^{+0.2}_{-0.2}$&$3.0^{+1.4}_{-1.2}$&\multicolumn{2}{c}{-}&0.019&0.05\\
$f_0(1710)\to\rho^0\gamma$&42 &94& 705&24&$55^{+16}_{-14}$&$410^{+200}_{-160}$&$100\pm40$&-&-&24\\
$f_0(1710)\to\omega\gamma$&-& - &-&-&-&-&$3.3\pm1.2$&-&-&82\\
$f_0(1710)\to\phi\gamma$&800 &718& 78&450&$400^{+20}_{-20}$&$36^{+17}_{-14}$&$15\pm5$&-&-&94\\
\hline\hline
    \end{tabular} 
\end{table*}

\subsection{Radiative decay widths of $f_0(1370)$ and $f_0(1710)$}
Now let us turn our attention to the $f_0(1370)$ and $f_0(1710)$ mesons. In
table \ref{table:f0comparison} we compare our results for the radiative decay widths of the
$f_0(1370)$ and $f_0(1710)$ obtained by the coupled channel model
with the predictions of other theoretical approaches, including the
nonrelativistic quark model (NRQM)~\cite{Close:2002sf}, the
light-front quark model (LFQM)~\cite{DeWitt:2003rs}, the
calculation of Nagahiro et al.~\cite{Nagahiro:2008bn}, and the chiral approach~\cite{Giacosa:2005zt}. In the NRQM
and LFQM calculations three numbers are given for each decay channel
depending on whether the glueball mass used in the
calculation is smaller than the $n\bar{n}$ mass (Light), between
the $n\bar{n}$ and $s\bar{s}$ masses (Medium), or larger than the $s\bar{s}$ mass (Heavy)~\cite{Close:2002sf,DeWitt:2003rs}.

 First we note that
for the $f_0(1370)$ our predicted two-photon decay width is more
consistent with the LFQM result in the light glueball scenario,
while the $\rho\gamma$ decay width lies closer to the LFQM result in the heavy glueball scenario.  Furthermore, the $\phi\gamma$ decay
width in our model is an order of magnitude smaller than that in the
LFQM.

For the $f_0(1710)$, the LFQM two-photon decay width is larger than
the current experimental limit (see Table \ref{table:res1}). On the other hand, our $\rho^0\gamma$
decay width is more consistent with the LFQM in the light gluon scenario while
the $\phi\gamma$ decay width is more consistent with that of the LFQM in the heavy gluon
scenario. Similar to the $f_0(1370)$ case, here further experimental
data are needed to clarify the situation.

Furthermore, we notice that the NRQM and the LFQM in the light and medium glueball mass scenarios
and our present study all predict that  $\Gamma_{\rho\gamma}\gg\Gamma_{\phi\gamma}$ for the $f_0(1370)$ while
 $\Gamma_{\rho\gamma}\ll\Gamma_{\phi\gamma}$ for the $f_0(1710)$. On the other hand, the NRQM and LFQM in the
heavy glueball scenario predict $\Gamma_{\rho\gamma}\gg\Gamma_{\phi\gamma}$ for the $f_0(1710)$. Therefore,
an experimental measurement of the ratio of $\Gamma_{f_0(1710)\rightarrow\rho\gamma}/\Gamma_{f_0(1710)\rightarrow\phi\gamma}$
not only will distinguish between the quark-model picture and the dynamical picture, but also will put a constraint on
the mass of a possible glueball in this mass region.

The chiral approach in Ref.~\cite{Giacosa:2005zt}
delivers smaller values for the two-photon decay rates of the $f_0(1370)$ and $f_0(1710)$.
 However, the ratio $\Gamma(f_0(1370)\to \gamma\gamma)/\Gamma(f_0(1710)\to \gamma\gamma)\approx18.4$
lies much closer to our prediction $\Gamma(f_0(1370)\to \gamma\gamma)/\Gamma(f_0(1710)\to \gamma\gamma)\approx26.2$ than
the LFQM results which range between 1.7--3.0.

The work of Nagahiro et al.~\cite{Nagahiro:2008bn} evaluates the
contribution from loops of $K\bar{K}$ ($\pi\pi$) using a phenomenological scalar coupling of the
$f_0(1710)$ ($f_0(1370)$) to $K\bar{K}$ ($\pi\pi$). From the new perspective
on these states we have after the work of Ref.~\cite{Geng:2008gx}, the scalar coupling may not be justified.
One rather has the $f_0(1710)$ coupling to $K^*\bar{K}^*$ while the coupling of the $K\bar{K}$ channel only occurs indirectly through the further decay $K^*\rightarrow K\pi$ and $\bar{K}^*\rightarrow\bar{K}\pi$,
with $\pi$ going into an internal propagator. As found in Ref.~\cite{Molina:2008jw}, loops containing
these $\pi$ propagators only lead to small contributions compared to leading terms including vector mesons (four-vector contact
and t(u)-channel vector exchange).

Experimentally, there is a further piece of information on the
$f_0(1710)$ that is relevant to the present study. From the $J/\psi$
decay branching ratios to $\gamma\omega\omega$ and $\gamma
K\bar{K}$, one can deduce \cite{Amsler:2008zzb}
\begin {equation}
\frac{\Gamma(f_0(1710)\to \omega\omega)}{\Gamma(f_0(1710)\to K\bar
K)}=\frac{Br(J/\psi\to \gamma f_0(1710)\to \gamma
\omega\omega)}{Br(J/\psi\to \gamma f_0(1710)\to \gamma K\bar
K)}=\frac{(3.1\pm
1.0)\times10^{-4}}{(8.5^{+1.2}_{-0.9})\times10^{-4}}=0.365^{+0.156}_{-0.169}\,.
\end{equation}
In the same way as we obtain the two-photon decay widths, we can
also calculate the two-vector-meson decay width of the dynamically generated resonances. For the $f_0(1710)$,
its decay width to $\omega\omega$ is found to be
\begin{align*}
\Gamma(f_0(1710)\to\omega\omega)&=15.2 \mbox{ MeV}.
 \end{align*}
 Using $\Gamma_\mathrm{total}(f_0(1710))=133$ MeV,
 already derived in Ref.~\cite{Geng:2009gb}, and the ratio $\frac{\Gamma(K\bar
K)}{\Gamma_\mathrm{total}(f_0(1710))}\approx 55$ \% also given in
Ref.~\cite{Geng:2009gb}, one obtains the following branching ratio
\begin {eqnarray}
\frac{\Gamma(f_0(1710)\to \omega\omega)}{\Gamma(f_0(1710)\to K\bar
K)}&=&0.21\, ,
\end{eqnarray}
which lies within the experimental bound, although close to the lower limit.

\subsection{Radiative decay widths of the $K^*_2(1430)$}
The radiative decay widths of the $K^*_2(1430)$ calculated in the present work
are compared with those calculated in the covariant oscillator quark model (COQM)~\cite{Ishida:1988uw} in
Table \ref{table:k2comparison}. We notice that the results from these two approaches differ by a factor of 10. However, there is one thing
in common, i.e., both predict a much larger $\Gamma_{K^{*0}\gamma}$  than the $\Gamma_{K^{*+}\gamma}$. More specifically,
in the COQM $\Gamma_{K^{*0}\gamma}/\Gamma_{K^{*+}\gamma}\approx 3$, while in our model this ratio is $\approx 4$.

At present there is no experimental measurement of these decay modes. On the other hand,
the $K^*_2(1430)\rightarrow K^+\gamma$ and $K^*_2(1430)\rightarrow K^0\gamma$ decay rates have been measured.
According to PDG~\cite{Amsler:2008zzb}, $\Gamma_{K^+\gamma}=241\pm50$ keV and $\Gamma_{K^0\gamma}<5.4$ keV.
Comparing these decay rates with those shown in Table~\ref{table:k2comparison}, one immediately notices that
the $\Gamma_{K^{*+}\gamma}$ in the dynamical model is of similar order as the $\Gamma_{K^+\gamma}$ despite reduced phase space
in the former decay, which is of course closely
related with the fact that the $K^*_2(1430)$ is built out of the coupled channel interaction between the $\rho K^*$, $\omega K^*$,
and $\phi K^*$ components in the dynamical model. Furthermore, both the COQM and our dynamical model predict
$\Gamma_{K^{*0}\gamma}\gg\Gamma_{K^{*+}\gamma}$, which is opposite to the decays into a kaon plus a photon where
$\Gamma_{K^+\gamma}\gg\Gamma_{K^0\gamma}$.
An experimental measurement of those decays would be very interesting and will certainly help
distinguish the two different pictures of the $K^*_2(1430)$.

\begin{table*}[t]
      \renewcommand{\arraystretch}{1.6}
     \setlength{\tabcolsep}{0.4cm}
     \centering
     \caption{Radiative decay widths of the $K^*_2(1430)$ (in keV) obtained in the present work in comparison with
those obtained in the covariant oscillator quark model (COQM)~\cite{Ishida:1988uw}.\label{table:k2comparison}}
     \begin{tabular}{ccc}
     \hline\hline
&COQM~\cite{Ishida:1988uw}&Present work\\\hline
$K^{\ast +}_2(1430)\to K^{\ast +} \gamma$& 38  &261\\
$K^{\ast 0}_2(1430)\to K^{\ast0}\gamma$ &109  & 1056\\
\hline\hline
    \end{tabular} 
\end{table*}

\section{Summary and conclusions}
We have calculated the radiative decay widths ($\gamma\gamma$ and $V\gamma$) of
the $f_2(1270)$, $f_0(1370)$, $f'_2(1525)$, $f_0(1710)$, $K^*_2(1430)$, and four other states
that appear dynamically from vector meson-vector meson interaction in a unitary approach.
Within this approach, due to the peculiarities of the hidden-gauge Lagrangians and the assumption that
these resonances are mainly formed by vector meson-vector meson interaction, one can factorize the
radiative decay process into  a strong part and an electromagnetic part. This way, the calculation is
greatly simplified and does not induce loop calculations. The obtained results are found to
be consistent with existing data within theoretical and experimental uncertainties.

When data are not available, we have compared our predictions with those obtained in other approaches.
In particular, we have identified the relevant pattern of decay rates predicted by different theoretical models and
found them quite distinct. For instance, the $\Gamma(f_2(1270)\rightarrow\rho\gamma)/\Gamma(f_2(1270)\rightarrow\omega\gamma)$
ratio is quite different in the dynamical model from those in the TMD\&VMD model and the COQM model.
The $\Gamma(f'_2(1525)\rightarrow\phi\gamma)/\Gamma(f'_2(1525)\rightarrow\rho\gamma)$ ratio in the COQM model
is also distinctly different from that in the dynamical model. A measurement of the $f_0(1370)$/$f_0(1710)$ decay rates into
$\rho\gamma$ and $\phi\gamma$ could be used not only to distinguish between the quark model (NRQM and LFQM) picture and the dynamical
picture but also to put a constraint on the mass of a possible glueball (in the $q\bar{q}$-g mixing scheme of the NRQM and LFQM).
For the $K^*_2(1430)$, as we have discussed, a measurement of its $K^{*+}(K^{*0})\gamma$ decay mode will definitely be able
to determine to what extent the dynamical picture is correct.

It is necessary to stress that the QCD dynamics is much richer than
that contained in our unitary approach. It is, therefore, not too
surprising to us that sometimes agreement with data is not perfect,
but the model delivers at least a qualitative insight into the decay
pattern. However, up to now the dynamical picture of the
$f_0(1370)$, $f_2(1270)$, $f'_2(1525)$, $f_0(1710)$, and
$K^*_2(1430)$ has been tested in a number of scenarios, including in
the $J/\psi\rightarrow VT$ decays~\cite{MartinezTorres:2009uk},
$J/\psi\rightarrow \gamma T$ decays~\cite{Geng:2009iw}, in their
strong decay modes~\cite{Geng:2009gb}, and in their two-photon decay
modes, as shown in Ref.~\cite{Nagahiro:2008um} and in the present
work. It will be interesting to see what comes out in their one
photon-one vector meson decay modes. Given their distinct pattern in
different theoretical models, an experimental measurement of some of
the decay modes will be very suggestive of the nature of these
resonances. Such measurements in principle could be carried out by
PANDA at FAIR or BESIII at BEPCII.

\section{Acknowledgements}
 L.S.G. acknowledges support from the MICINN in the Program ``Juan de la Cierva.''
 This work is partly supported by DGICYT Contract No. FIS2006-03438, the EU Integrated
Infrastructure Initiative Hadron Physics Project under contract
RII3-CT-2004-506078 and the DFG under contract No. GRK683.


\begin{thebibliography}{99}
\bibitem{Amsler:2008zzb}
  C.~Amsler {\it et al.}  [Particle Data Group],
  Phys.\ Lett.\  B {\bf 667}, 1 (2008).

\bibitem{Oller:1997ti}
  J.~A.~Oller and E.~Oset,
  Nucl.\ Phys.\  A {\bf 620}, 438 (1997)
  [Erratum-ibid.\  A {\bf 652}, 407 (1999)].

\bibitem{Kaiser:1998fi}
  N.~Kaiser,
  Eur.\ Phys.\ J.\  A {\bf 3}, 307 (1998).



\bibitem{Markushin:2000a}
  V.~E.~Markushin,
  Eur.\ Phys.\ J.\  A {\bf 8}, 389 (2000).

\bibitem{Dobado:1997a}
  A.~Dobado and J.~R.~Pelaez,
  Phys.\ Rev.\  D {\bf 56}, 3057 (1997).

  \bibitem{ramonet}
  J.~A.~Oller, E.~Oset and J.~R.~Pelaez,
  Phys.\ Rev.\  D {\bf 59}, 074001 (1999)
  [Erratum-ibid.\  D {\bf 60}, 099906 (1999);  Erratum-ibid. D75, 099903 (2007)].



\bibitem{Kaiser:1995eg}
  N.~Kaiser, P.~B.~Siegel and W.~Weise,
  Nucl.\ Phys.\  A {\bf 594}, 325 (1995).

  \bibitem{Oset:1998a}
  E.~Oset and A.~Ramos,
  Nucl.\ Phys.\  A {\bf 635}, 99 (1998).

  \bibitem{ollerulf}
  J.~A.~Oller and U.~G.~Meissner,
  Phys.\ Lett.\  B {\bf 500}, 263 (2001).

  \bibitem{Recio:2004a}
  C.~Garcia-Recio, M.~F.~M.~Lutz and J.~Nieves,
  Phys.\ Lett.\  B {\bf 582}, 49 (2004).

  \bibitem{Jido:2003a}
  D.~Jido, J.~A.~Oller, E.~Oset, A.~Ramos and U.~G.~Meissner,
  Nucl.\ Phys.\  A {\bf 725}, 181 (2003).

  \bibitem{Recio:2006a}
  C.~Garcia-Recio, J.~Nieves and L.~L.~Salcedo,
  Phys.\ Rev.\  D {\bf 74}, 034025 (2006).

  \bibitem{Hyodo:2003a}
  T.~Hyodo, S.~I.~Nam, D.~Jido and A.~Hosaka,
  Phys.\ Rev.\  C {\bf 68}, 018201 (2003).

  \bibitem{Borasoy:2005a}
  B.~Borasoy, R.~Nissler and W.~Weise,
  Eur.\ Phys.\ J.\  A {\bf 25}, 79 (2005).

  \bibitem{Oller:2005a}
  J.~A.~Oller, J.~Prades and M.~Verbeni,
  Phys.\ Rev.\ Lett.\  {\bf 95}, 172502 (2005).


\bibitem{Borasoy:2006a}
  B.~Borasoy, U.~G.~Meissner and R.~Nissler,
  Phys.\ Rev.\  C {\bf 74}, 055201 (2006).



\bibitem{Kolomeitsev:2003ac}
  E.~E.~Kolomeitsev and M.~F.~M.~Lutz,
  Phys.\ Lett.\  B {\bf 582}, 39 (2004);
  J.~Hofmann and M.~F.~M.~Lutz,
  Nucl.\ Phys.\  A {\bf 733}, 142 (2004);
  F.~K.~Guo, P.~N.~Shen, H.~C.~Chiang and R.~G.~Ping,
  Phys.\ Lett.\  B {\bf 641}, 278 (2006);
  D.~Gamermann, E.~Oset, D.~Strottman and M.~J.~Vicente Vacas,
  Phys.\ Rev.\  D {\bf 76}, 074016 (2007).



\bibitem{MartinezTorres:2007sr}
  A.~Martinez Torres, K.~P.~Khemchandani and E.~Oset,
  Phys.\ Rev.\  C {\bf 77}, 042203 (2008);
  A.~Martinez Torres, K.~P.~Khemchandani, L.~S.~Geng, M.~Napsuciale and E.~Oset,
  Phys.\ Rev.\  D {\bf 78}, 074031 (2008).


\bibitem{Molina:2008jw}
  R.~Molina, D.~Nicmorus and E.~Oset,
  Phys.\ Rev.\  D {\bf 78}, 114018 (2008).

 \bibitem{Gonzalez:2008pv}
  P.~Gonzalez, E.~Oset and J.~Vijande,
  Phys.\ Rev.\  C {\bf 79}, 025209 (2009).



\bibitem{Sarkar:2009kx}
  S.~Sarkar, B.~X.~Sun, E.~Oset and M.~J.~V.~Vacas,
  arXiv:0902.3150 [Eur. Phys. J. A (to be published)].


\bibitem{Molina:2009eb}
  R.~Molina, H.~Nagahiro, A.~Hosaka and E.~Oset,
  Phys.\ Rev.\  D {\bf 80}, 014025 (2009).


\bibitem{Oset:2009vf}
  E.~Oset and A.~Ramos,
  arXiv:0905.0973 [Eur. Phys. J. A (to be published)].


\bibitem{Geng:2008gx}
  L.~S.~Geng and E.~Oset,
   Phys.\ Rev.\  D {\bf 79}, 074009 (2009).

\bibitem{Geng:2009gb}
  L.~S.~Geng, E.~Oset, R.~Molina and D.~Nicmorus,
  arXiv:0905.0419.

\bibitem{Nagahiro:2008um}
  H.~Nagahiro, J.~Yamagata-Sekihara, E.~Oset, S.~Hirenzaki, and R.~Molina,
  Phys.\ Rev.\  D {\bf 79}, 114023 (2009).


\bibitem{MartinezTorres:2009uk}
  A.~Martinez Torres, L.~S.~Geng, L.~R.~Dai, B.~X.~Sun, E.~Oset and B.~S.~Zou,
  Phys.\ Lett.\  B {\bf 680}, 310 (2009).


\bibitem{Geng:2009iw}
  L.~S.~Geng, F.~K.~Guo, C.~Hanhart, R.~Molina, E.~Oset and B.~S.~Zou,
  arXiv:0910.5192 [Eur. Phys. J. A (to be published)].


\bibitem{Pennington:2008qa}
  M.~R.~Pennington,
  Nucl.\ Phys.\ Proc.\ Suppl.\  {\bf 181-182}, 251 (2008).

\bibitem{Amsler:2002ey}
  C.~Amsler,
  Phys.\ Lett.\  B {\bf 541}, 22 (2002).



\bibitem{Bando:1984ej}
  M.~Bando, T.~Kugo, S.~Uehara, K.~Yamawaki and T.~Yanagida,
  Phys.\ Rev.\ Lett.\  {\bf 54}, 1215 (1985);
  M.~Bando, T.~Kugo and K.~Yamawaki,
  Phys.\ Rept.\  {\bf 164}, 217 (1988).


\bibitem{tornqvist1}
 N.~A.~Tornqvist and M.~Roos,
 Phys.\ Rev.\ Lett.\  {\bf 76}, 1575 (1996).

\bibitem{tornqvist2}
 N.~A.~Tornqvist,
 Z.\ Phys.\  C {\bf 68}, 647 (1995).

\bibitem{beveren}
 E.~van Beveren, T.~A.~Rijken, K.~Metzger, C.~Dullemond, G.~Rupp and J.~E.~Ribeiro,
 Z.\ Phys.\  C {\bf 30}, 615 (1986).

\bibitem{boglione}
  M.~Boglione and M.~R.~Pennington,
  Phys.\ Rev.\  D {\bf 65}, 114010 (2002).

\bibitem{kaiser}
 N.~Kaiser, P.~B.~Siegel and W.~Weise,
 Phys.\ Lett.\  B {\bf 362}, 23 (1995).

\bibitem{inoue}
 T.~Inoue, E.~Oset and M.~J.~Vicente Vacas,
 Phys.\ Rev.\  C {\bf 65}, 035204 (2002).

\bibitem{vijande}
 A.~Valcarce and J.~Vijande,
 arXiv:0912.3080.


\bibitem{vijande2}
 J.~Vijande, A.~Valcarce and N.~Barnea,
 Phys.\ Rev.\  D {\bf 79}, 074010 (2009).




\bibitem{hyodo}
 T.~Hyodo, D.~Jido and A.~Hosaka,
 Phys.\ Rev.\  C {\bf 78}, 025203 (2008).


\bibitem{bugg}
 D.~V.~Bugg,
 arXiv:1001.1712.



\bibitem{jose1}
 C.~Hanhart, J.~R.~Pelaez and G.~Rios,
 Phys.\ Rev.\ Lett.\  {\bf 100}, 152001 (2008).

\bibitem{jose2}
 J.~R.~Pelaez and G.~Rios,
 arXiv:0905.4689.

\bibitem{giacosa1}
  F.~Giacosa,
  Phys.\ Rev.\  D {\bf 80}, 074028 (2009).

\bibitem{Li:2000zb}
  D.~M.~Li, H.~Yu and Q.~X.~Shen,
  J.\ Phys.\ G {\bf 27}, 807 (2001).
  
  
\bibitem{klempt}
E.~Klempt and A.~Zaitsev,
Phys.\ Rept.\  {\bf 454}, 1 (2007).

\bibitem{crede}
V.~Crede and C.~A.~Meyer,
Prog.\ Part.\ Nucl.\ Phys.\  {\bf 63}, 74 (2009).


\bibitem{isgur}
  S.~Godfrey and N.~Isgur,
  Phys.\ Rev.\  D {\bf 32}, 189 (1985).


\bibitem{Barnes:1996ff}
  T.~Barnes, F.~E.~Close, P.~R.~Page and E.~S.~Swanson,
  Phys.\ Rev.\  D {\bf 55}, 4157 (1997).

\bibitem{Barnes:2002mu}
  T.~Barnes, N.~Black and P.~R.~Page,
  Phys.\ Rev.\  D {\bf 68}, 054014 (2003).


  
\bibitem{Anisovich:2002im}
  A.~V.~Anisovich, V.~V.~Anisovich, M.~A.~Matveev and V.~A.~Nikonov,
  Phys.\ Atom.\ Nucl.\  {\bf 66}, 914 (2003)
  [Yad.\ Fiz.\  {\bf 66}, 946 (2003)].








\bibitem{gonzalez}
P. Gonzalez, private communication.

\bibitem{Nacher:1999ni}
  J.~C.~Nacher, E.~Oset, H.~Toki and A.~Ramos,
  Phys.\ Lett.\  B {\bf 461}, 299 (1999).


\bibitem{Borasoy:2007ku}
  B.~Borasoy, P.~C.~Bruns, U.~G.~Meissner and R.~Nissler,
  Eur.\ Phys.\ J.\  A {\bf 34}, 161 (2007).

\bibitem{Adachi:1989dd}
  I.~Adachi {\it et al.}  [TOPAZ Collaboration],
  Phys.\ Lett.\  B {\bf 234}, 185 (1990).

\bibitem{Morgan:1990kw}
  D.~Morgan and M.~R.~Pennington,
  Z.\ Phys.\  C {\bf 48}, 623 (1990).
  
\bibitem{Behrend:1988hw}
  H.~J.~Behrend {\it et al.}  [CELLO Collaboration],
  Z.\ Phys.\  C {\bf 43}, 91 (1989).

\bibitem{Longacre:1986fh}
  R.~S.~Longacre {\it et al.},
  Phys.\ Lett.\  B {\bf 177}, 223 (1986).




\bibitem{Ishida:1988uw}
  S.~Ishida, K.~Yamada and M.~Oda,
  Phys.\ Rev.\  D {\bf 40}, 1497 (1989).

\bibitem{Suzuki:1993zs}
  M.~Suzuki,
  Phys.\ Rev.\  D {\bf 47}, 1043 (1993).

\bibitem{Katz:2005ir}
  E.~Katz, A.~Lewandowski and M.~D.~Schwartz,
  Phys.\ Rev.\  D {\bf 74}, 086004 (2006).

\bibitem{Oh:2003aw}
  Y.~s.~Oh and T.~S.~H.~Lee,
  Phys.\ Rev.\  C {\bf 69}, 025201 (2004).

\bibitem{Close:2002sf}
  F.~E.~Close, A.~Donnachie and Yu.~S.~Kalashnikova,
  Phys.\ Rev.\  D {\bf 67}, 074031 (2003).

\bibitem{Boyer:1990vu}
  J.~Boyer {\it et al.},
  Phys.\ Rev.\  D {\bf 42}, 1350 (1990).

\bibitem{Giacosa:2005bw}
  F.~Giacosa, T.~Gutsche, V.~E.~Lyubovitskij and A.~Faessler,
  Phys.\ Rev.\  D {\bf 72}, 114021 (2005).






\bibitem{DeWitt:2003rs}
  M.~A.~DeWitt, H.~M.~Choi and C.~R.~Ji,
  Phys.\ Rev.\  D {\bf 68}, 054026 (2003).

\bibitem{Nagahiro:2008bn}
  H.~Nagahiro, L.~Roca, E.~Oset and B.~S.~Zou,
  Phys.\ Rev.\  D {\bf 78}, 014012 (2008).



\bibitem{Giacosa:2005zt}
  F.~Giacosa, T.~Gutsche, V.~E.~Lyubovitskij and A.~Faessler,
  Phys.\ Rev.\  D {\bf 72}, 094006 (2005).



\end{thebibliography}
\end{document}